\documentclass[letterpaper, 10 pt, conference]{IEEEtran}  % Comment this line out if you need a4paper
\onecolumn

\IEEEoverridecommandlockouts
\usepackage{graphics,graphicx} % for pdf, bitmapped graphics files
\usepackage{epsfig} % for postscript graphics files

\usepackage{amsmath}
\usepackage{amssymb}
\usepackage{tikz}
\usepackage{amsthm}

\usepackage{comment}
\usepackage[export]{adjustbox}
\usepackage{mathcomSTEP}
\usepackage{subcaption}
\usepackage[utf8]{inputenc}

\usepackage{tikz}
\usetikzlibrary{positioning}

\newtheorem{prop}[thm]{Proposition}
\newtheorem{rem}[thm]{Remark}

\title{\LARGE \bf
Indirect Rate-Distortion Function of a Binary i.i.d Source}

\author{Alon Kipnis$^{*}$, Stefano Rini$^{\dagger}$ and  Andrea J. Goldsmith$^{*}$
\thanks{$^{*}$Department of Electrical Engineering, Stanford University, USA.  }

\thanks{$^{\dagger}$Department of Electrical and Computer Engineering, National Chiao Tung University (NCTU), Taiwan.}}
\begin{document}
\graphicspath{{../Figures/}}

\tikzstyle{int}=[draw, fill=blue!10, minimum height = 1cm, minimum width=1.5cm,thick ]
\tikzstyle{sum}=[circle, fill=blue!10, draw=black,line width=1pt,minimum size = 0.3cm, thick ]

\maketitle
\thispagestyle{empty}
\pagestyle{empty}

%%%%%%%%%%%%%%%%%%%%%%%%%%%%%%%%%%%%%%%%%%%%%%%%%%%%%%%%%%%%%%%%%%%%%%%%%%%%%%%%
\begin{abstract}
The indirect source-coding problem in which a Bernoulli process is compressed in a lossy manner from its noisy observations is considered.
These noisy observations are obtained by passing the source sequence through a binary symmetric channel so that the channel crossover probability controls the amount of information available about the source realization at the encoder. 
We use classic results in rate-distortion theory to compute an expression of the rate-distortion function for this model, where the Bernoulli source is not necessarily symmetric. The indirect rate-distortion function is given in terms of a solution to a simple equation. In addition, we derive an upper bound on the indirect rate-distortion function which is given in a closed.
These expressions capture precisely the expected behavior that the noisier the observations, the smaller the return from increasing bit-rate to reduce distortion.
\end{abstract}


%%%%%%%%%%%%%%%%%%%%%%%%%%%%%%%%%%%%%%%%%%%%%%%%%%%%%%%%%%%%%%%%%%%%%%%%%%%%%%%%

\section{INTRODUCTION}
\label{sec:Intro}
The optimal trade-off between bit-rate and average distortion in the representation of an information source is given by the Rate-Distortion Function (RDF):
%
%the RDF provides the minimum rate necessary to describe a source when its reconstruction is required to lye to within a prescribed fidelity from the original sequence.
the RDF provides the minimum rate necessary to describe a source when its reconstruction is allowed to be to within a given average distortion from the original sequence.
%
%communication over a rate-limited link under a fidelity criterion by specifying a lower bound on the minimal code rate such that the average distortion is bounded from by a specified value,
%A natural extension of this source coding problem and the associated rate-distortion trade-off is the case in which the encoder cannot observe the source directly.
A natural extension of this source coding problem is the scenario in which the encoder cannot observe the source directly but obtains only noisy observations.
This could be due to a number of phenomena such as environmental noise, finite precision quantization and sub-sampling \cite{Kipnis2014}.
In this setup, the encoder is required to describe the source from another process statistically correlated with the source itself: this problem is known as \emph{indirect} or \emph{remote} source coding \cite[Sec. 3.5]{berger1971rate}.

 \par
An interesting motivation for the indirect source coding problem arises in centralized sensing networks in which each sensor in the network is required to transmit its observation to a remote processing unit.
Restrictions on the computational complexity and power consumption of the sensors make local processing infeasible and thus the uncompressed data has to be communicated over the network.
%
%The data acq from each sensor can be a noisy observation of an underlying information source.
The communication toward the central unit introduces noise in the sensors' observations and the compression rate of the data acquired at the central node is determined by the indirect RDF.

The general structure of an indirect source coding problem is depicted in Figure~\ref{fig:indirect_system_model}: the source  process, $X^n$, is passed through the noisy channel $P_{Y|X}^n$ to obtain the signal $Y^n$. The encoder compresses the sequence $Y^n$ at rate $R$ and the compressed observation is provided noiselessly to the decoder.
The receiver produces the sequence $\Xh^n$ which is a reconstruction of the original signal $X^n$ to within a prescribed average distortion.
\par
While in the direct source coding problem the RDF describes the optimal trade-off between the code rate $R$ and distortion $D$, another quantity of merit in the indirect problem is the channel $P_{Y|X}$. By characterizing the trade-off in the indirect problem, namely by an \emph{indirect RDF}, it is possible to study the effect of the channel quality on the optimal rate-distortion trade-off. For instance, it is of interest to characterize the amount of additional code-rate needed to maintain a fixed distortion level as the observations become noisier.
\par
%
%The fidelity criterion is measured between the reconstruction sequence $\Xh$ and $X$, while the code is applied to the observable process $Y$. The `channel' $\mathrm P_{Y|X}$ between $X$ and $Y$ describes incomplete information of the source $X$ available at the encoder. \\

It has long been noticed \cite{berger1971rate,1056251,1054469} that an indirect source coding problem can be reduced to a standard source coding problem by the following argument:
it is possible to consider the observable process $Y^n$ as the source in the standard source coding problem by amending the fidelity criterion to capture the distance between the reconstructed symbol $\Xh^n$ and all possible realizations of the original source realization $X^n$ weighed according to the probability of their appearance given $Y^n$.
%
%consider the observable process $Y$ as the source; amend the fidelity criterion to capture the distance between the reconstruction symbol $\Xh$ and all possible realizations of the original source symbols weighed according to their corresponding probabilities given $y$.
%
A particularly intuitive form of this observation appears in the case of a quadratic distortion, where the amended fidelity criterion can be decomposed as the sum of two terms:
 (i) the mean squared error (MSE) estimation of the source from its observation plus (ii) the error in describing the MSE estimate under a rate-limited description \cite{1054469}.
This separation allows one to obtain the closed form expression of the indirect RDF in the Gaussian source, quadratic distortion and additive Gaussian noise case \cite{1057738, Kipnis2014}.
\par
While, in general, similar separation results for other models do not exist, it may still be possible to solve the direct problem using the amended distortion measure.
This approach is explored in this paper for the important case of a binary i.i.d source, bit flipping noise and the Hamming distortion.

%\subsection*{Related Works}
\emph{\bf Related Work:}
The source coding problem  was first introduced by Shannon in \cite{Shannon1948} while he provided the first of the source coding theorem in \cite{shannon1959coding}.
Indirect rate-distortion problem was first introduced by Dobrushin and Tsybakov in \cite{1057738}.
The authors of \cite{1057738} derived a closed form solution for the indirect RDF in the Gaussian stationary case and, implicitly, showed an equivalence of the indirect problem to a direct source coding problem with an amended fidelity criterion.
Wolf and Ziv \cite{1054469} showed that, in the case of a quadratic distortion, the new fidelity criterion identified in \cite{1056251}  decomposes into the sum of two terms,  only one of which depends on the source coding rate $R$. Wolf and Ziv also computed the indirect RDF (iDRF) in various cases include the case of a Bernoulli source observed through a binary symmetric channel under quadratic distortion. Since the quadratic distortion of a binary sequence is not larger than its Hamming distance, their result provides a lower bound on the iRDF of the same source under the Hamming distance considered in this work. Berger \cite{berger1971rate} noted the equivalence of the indirect problem to a modified direct problem with a new fidelity criterion, and gave an interpretation of the new fidelity criterion as the conditional expectation of the the original distortion measure given the source and noise realizations.\par
In the special case of a Bernoulli observed through a binary channel, the computation of the iRDF is greatly simplified when the source is symmetric $\mathbb P(X_n=1)=1/2$ \cite[Exc. 3.8]{berger1971rate}. In our setting where the observation are given by a binary symmetric channel, this iDRF is given by
\begin{equation} \label{eq:result 1}
R_{X|Y}(D) = \begin{cases}
1-h\left(\frac{D-p}{1-2p} \right) & p<D<1/2\\
0 & D\geq 1/2,
\end{cases}
\end{equation}
where $h(x)$ is the binary entropy function and  $p<1/2$ and $\po=1-p$ (the case $p>1/2$ can be treated in a similar fashion). The symmetric case can also be obtained as a special case from \cite{yamamoto1980source} where indirect versions of the multiterminal setting of Slepian-Wolf %\cite{1055037}
and Wyner-Ziv problems
%\cite{1055508} 
were considered. 

%Witsenhausen \cite{1056251} noted the equivalence of the indirect problem to a modified direct problem with a new fidelity criterion, and extended the equivalence to the case in which side information is available at the decoder.

%\subsection*{Contribution}
%
\emph{\bf Contributions:}
We derive an expression for the iRDF of a Bernoulli process $X^n$ with $\mathbb P(X_n=1)=\al$ given its observation $Y^n$ through a binary symmetric channel with crossover probability $p$ for the general case of $\al \in [0,1/2)$. 
%In the symmetric case where $\mathbb P(X_n=1)=\al=1/2$, this iRDF can be obtained from \cite[Exc. 3.8]{berger1971rate} or \cite[Eq. 36]{yamamoto1980source} to be:
%\begin{equation} \label{eq:result 1}
%R_{X|Y}(D) = \begin{cases}
%1-h\left(\frac{D-p}{1-2p} \right) & p<D<1/2\\
%0 & D\geq 1/2,
%\end{cases}
%\end{equation}
%where $h(x)$ is the binary entropy function and  $p<1/2$ and $\po=1-p$ (the case $p>1/2$ can be treated in a similar fashion). 
%
This iRDF $R_{X|Y}(D)$ is obtained by finding the unique root to an equation whose parameters are determined by $\al$, $p$ and $D$. Additionally, we show that an upper bound on $R_{X|Y}(D)$ expressed as (for $p\leq 0.5$)
\begin{equation} \label{eq:result 2}
\overline{R}_{X|Y}(D)=
 \begin{cases}
h\left(\al \star p\right) -h\left(\frac{D-p}{1-2p} \right) & p\leq D \leq 1/2\\
0 & D\geq \al,
\end{cases}
\end{equation}
where  $\al \star  p \triangleq p\bar{\al} + \al \bar{p}$ with equality if and only if $\al = 1/2$, in which case $R_{X|Y}(D)=\overline{R}_{X|Y}(D)$ for all $D$.
%$\eqref{eq:result 1}=\eqref{eq:result 2}$.
\smallskip

The rest of this paper is organized as follows:
the indirect source coding problem and the relevant background literature  are introduced in Sec. \ref{sec:pre}.
The main results are derived in Sec. \ref{sec:Results}.
Finally, Sec. \ref{sec:Conclusions} concludes the paper.

%\smallskip
%\noindent
%\underline{Only sketches of proofs appear in this manuscript: full proofs}
%\underline{can be found in an extended version available online \cite{Kipnis2015ITW}.}
%{\color{red} I am ok with that, but we need to write the extended version as well .}
%SR we can split the effort! no sweat!

\section{Problem Statement }

\label{sec:pre}
\begin{figure}
\centering
\begin{tikzpicture}[node distance=2.5cm,auto,>=latex]
  \node at (0,0) (source) {$X^n$};
   \node [int] (Pxy) [right of = source, node distance = 1.8 cm]{$\mathrm{P}_{Y^n|X^n}$};
    \node [int] (enc) [right of = Pxy]{$\mathrm{Enc}$};
    \node [int] (dec) [right of=enc, node distance = 2.3cm] {$\mathrm{Dec}$};
    \node [right] (dest) [right of=dec,node distance =1.8 cm]{$\hat{X^n}$};
   \draw[->,line width=2pt] (enc) --node[above] {$R$} (dec) ;
   \draw[->,line width=2pt] (dec) -- (dest);
   \draw[->,line width=2pt] (source) -- (Pxy);
   \draw[->,line width=2pt] (Pxy) -- node[above] {$Y^n$}(enc);
 \end{tikzpicture}
\caption{\label{fig:indirect_system_model} Indirect source coding model}
%\vspace{-.5 cm }
\end{figure}
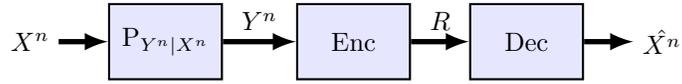

%\subsection{Definition of the indirect RDF}
%\underline{\bf Definition of the indirect RDF}
We consider the indirect source coding problem depicted in Fig. \ref{fig:indirect_system_model}: 
an encoder observes the discrete time process $X^n$ through the noisy channel $P_{Y^n|X^n}$ and produces a sequence of coded symbols at rate $R$.
From this sequence of coded symbols, the decoder produces a reconstructed sequence $\Xh^n$ which must be to within maximum average distortion from $X^n$ for a prescribed fidelity criterion.

More specifically, given a source sequence $X^n \triangleq \left\{X_k, \ k=1,2 \ldots n \right\}$ with alphabet $\Xcal^n$, the encoder is provided with the sequence $Y^n$ with alphabet $\Ycal^n$  obtained from $X^n$ through the  channel  $P_{Y^n|X^n}(Y^n|X^n)$ and maps this sequence unto the set $\lcb 1 \ldots 2^{\lfloor Rn \rfloor } \rcb$ through the mapping
%
%and is limited to use $n$-block codes, i.e., the encoding-decoding rule for a code of block-length $n\in \mathbb N$ is given by a mapping
\ea{
 W(Y^n): \ \  \Ycal^n \rightarrow \left\{1 \ldots 2^{\lfloor Rn \rfloor }\right\}.
\label{eq:encoding mapping}
}
The value $W(Y^n)$ is noiselessly communicated to the receiver which, in turns,  produces the sequence $\Xh^n$ with alphabet $\widehat{\Xcal^n}$ through the mapping
\ea{
\Xh^n(W): \ \ \lcb1 \ldots 2^{\lfloor Rn \rfloor } \rcb \rightarrow  \widehat{\Xcal}^n.
 \label{eq:decoding mapping}
}
%
%
%
%The encoding and decoding mappings are subject to a rate and a distortion constraint,
%%
%
The sequence $\Xh^n$ must be to within a distortion $D$ from $X^n$ for some chosen fidelity criterion $d_n(x^n,\xh^n)$
%
%The average distortion between $X^n$ and the reconstruction $\Xh^n$ must be less than a given threshold $D$
%so that average distortion between $X^n$ and $\Xh^n$ is 
which is measured with the per-letter distortion function $d\left(x_i,\xh_i \right)$, as
%The fidelity criterion between a source realization $X^n=x^n$ and its reconstruction $\Xh=\xh^n$ is measured as
\ea{
d_n(x^n,\xh^n) \triangleq \sum_{i=1}^n d\left(x_i,\xh_i \right),
\label{eq:average distortion}
}
for some real-valued, bounded function $d(\cdot,\cdot)$.

The operational indirect RDF $\tilde{R}_{X|Y}(D)$ is defined as the minimal rate $R$ in \eqref{eq:encoding mapping} and \eqref{eq:decoding mapping} such that the average distortion between $X^n$ and $\Xh^n$ in \eqref{eq:average distortion} does not exceed $D$, as the block-length $n$ goes to infinity. \\

The indirect (Shannon's) RDF (iRDF) for the channel $P_{Y^n|X^n}$ is defined as
\[
R_{X|Y}(D) =\liminf_{n\rightarrow \infty} R_n(D),
\]
where
\[
R_n(D) = \inf \frac{1}{n} I\left( Y^n; \hat{X}^n \right) \leq R,
\]
and the infimum is taken over all mappings $Y^n \rightarrow \hat{X}^n = \eqref{eq:encoding mapping} \circ  \eqref{eq:decoding mapping}$  such that the average distortion between $X^n$ and $\hat{X}^n$ is at most $D$. \\

The customary source coding problem \cite{shannon1959coding}, also \emph{direct} source coding problem, is obtained from the indirect source coding problem by simply letting $Y^N=X^N$. 
It is noted in \cite{1056251} that the problem of finding the operational indirect source coding rate $\Rt_{X|Y}(D)$ can be reduced to a direct source
coding problem for the observable process $Y^n$ and a different distortion measure $\dh(\cdot,\cdot)$ defined as
\begin{equation} \label{eq:d_bar_def}
\dh_n ( y^n,\xh^n) \triangleq \mathbb E\left[d_n( X^n,\hat{ x}^n)|Y^n=  y^n \right].
\end{equation}
Note that $\dh(\cdot,\cdot)$ depends only on $d(\cdot,\cdot)$ and $P_{Y^n|X^n}$, which are determined by the structure of the original indirect rate distortion problem.

Since
\[
\mathbb E \lsb d_n\left(X^n,\Xh^n \right) \rsb = \mathbb E \lsb \dh_n  \left(Y, \Xh \right) \rsb,
\]
it follows that $R_{X|Y}(D)$ equals the (direct) RDF $R_Y(D)$ of the process $Y^N$ under the fidelity criterion $\dh(\cdot,\cdot)$.
%
%Shannon Indirect RDF (iRDF) is defined as
%%In \cite{1057738} it was shown that
%\[
%R_{X|Y}^S(D) = \liminf_n \frac{1}{n} I\left( Y^n, \Xh^n \right),
%\]
Shannon's source coding theorem \cite{shannon1959coding} now implies
\begin{equation} \label{eq:indirect_source_coding_theorem}
\tilde{R}_{X|Y}(D) = R_Y(D) = R_{X|Y}(D).
\end{equation}
%The special case of \eqref{eq:indirect_source_coding_theorem} where the fidelity criterion is quadratic was derived in \cite{1054469}. \\

The reduction of the indirect source coding problem to a direct problem under $\hat{d}(\cdot,\cdot)$ also provides us with an approach to solve the indirect problem.
Namely, one can compute the direct distortion $\hat{d}(\cdot,\cdot)$ and compute the RDF for the source $Y^N$ under $\hat{d}(\cdot,\cdot)$.

%
%In the special case in which $P_{Y^n|X^n}$ factorizes as
%\ea{
%P_{Y^n|X^n}=\prod_{i=1}^n P_{Y|X}(y_i|x_i)
%\label{eq:memoryless}
%}
%for some $P_{Y|X}(y|x)$, convexity of the rate distortion function can be invoked to conclude that
%\ea{
%R_{X|Y}(D) \leq \min I(Y;X)
%\label{eq:memoryless RDF}
%}
%where the minimum is over all the mappings $Y^n\rightarrow \hat{X}^n $ such that the per-letter average amended distortion is at most $D$.
%%the indirect rate distortion function takes the

\subsection{Relevant results}
The computation of a direct RDF $R_{U}(D)$ of a source $U$ over a discrete alphabet $\mathcal U$ is performed by minimizing the mutual 
information over the  set of transition probabilities
\[
P(\hat{u}|u) \triangleq \mathbb P(\hat{U}=\hat{u}|U=u),
\]
under the constraint
\[
\sum_{u\in \mathcal U} \sum_{\hat{u}\in \hat{\mathcal U} }Q(u)P(\hat{u}|u)d(u,\hat{u}) \leq D,
\]
where $Q(u) \triangleq \mathbb P(U=u)$ and $d(\cdot,\cdot)$ is the per-letter distortion measure. 
This is equivalent to finding a stationary point to the Lagrangian
\begin{equation} \label{eq:lagrangian}
 L_0(r,\Pv)=\sum_{u,\hat{u}}  Q(u)P(\hat{u}|u) \left[ \log \frac{P(\hat{u}|u)}{\sum_{u,\hat{u}} Q(u)P(\hat{u}|u)}  +r(d(u,\hat{u})-D) \right]
\end{equation}
over the set of all transition probabilities. By introducing the constraint on the transition probabilities and using the Lagrange dual of \eqref{eq:lagrangian}, Gallager proved in \cite{gallager1970information} the theorem below.

\begin{thm} {\bf \cite[Thm. 9.4.1]{gallager1970information} }
\label{thm:gallager}
For a given source entropy $H(U)$ and a given distortion measure $d(\cdot,\cdot)$, let
\[
R_0(r, \Pv) \triangleq \sum_{u,\hat{u}} Q(u)P(\hat{u}|u)\left[\ln\frac{P(\hat{u}|u)}{\sum_{u} Q(u)P(\hat{u}|u)} + r d(u,\hat{u}) \right],
\]
then for any $r>0$,
\ea{\label{eq:gallager_bound}
\min_{\Pv} R_0(r,\Pv) = H(U) + \max_{\fv} \sum_u Q(u) \ln f_{u},
}
where the minimization in the LHS of \eqref{eq:gallager_bound} is over all transition probability functions 
$\Pv = \left\{ P(\hat{u}|u),\, u\in \mathcal U,\, \hat{u} \in \hat{\mathcal U} \right\}$, and the maximization in the RHS of \eqref{eq:gallager_bound}  is over all $\fv = \lcb f_u,\, u\in \mathcal U \rcb$ with non-negative components satisfying the constraints
\begin{equation} \label{eq:gallager_fk}
 \sum_u f_u e^{-r d(u,\hat{u} )} \leq 1,\quad \hat{u}\in \hat{\mathcal U}.
\end{equation}
Necessary and sufficient conditions on $\fv$ to achieve the maximum in \eqref{eq:gallager_bound} are the existence of a set of non-negative numbers $\left\{ w(\hat{u}),\,\hat{u}\in \hat{\mathcal U} \right\}$ satisfying
\begin{equation} \label{eq:gallager_w}
1 = \frac{f_u}{Q(u)} \sum_{\hat{u} \in \hat{\mathcal U}} w(\hat{u})e^{-r d(u,\hat{u})},
\end{equation}
and that \eqref{eq:gallager_fk} is satisfied with equality for each $\hat{u}$ with $w(\hat{u})>0$.
\end{thm}
It follows from \eqref{eq:lagrangian} that if the conditions for equality in Theorem~\ref{thm:gallager} hold, we have
\[
R_U(D) = \min_{\Pv} R_0(r,\Pv) = H(U) + \max_{\fv} \sum_u Q(u) \ln f_{u}.
\]
We refer to \cite{1266800} for a discussion of Theorem~\ref{thm:gallager} in the context of convex optimization theory as well as a geometric programming representation of this problem.

\begin{figure}
\centering
\begin{tikzpicture}[auto,>=latex]
  \node at (0,0) (source) {$X_n$};
   \node[right of = source, node distance = 2.5cm] (dest) {$Y_n$};
   \node[above of = source, node distance =0.5cm, xshift = 0.2] (s1){$1$};
  \node[below of = source, node distance =0.5cm, xshift = 0.2] (s0){$0$};
  \node[above of = dest, node distance =0.5cm,xshift = -0.2] (d1){$1$};
  \node[below of = dest, node distance =0.5cm,xshift = -0.2] (d0){$0$};
 \draw[->,line width = 1pt] (s0) -- node[below] {\scriptsize $1-p$} (d0);
   \draw[->,line width = 1pt] (s1)-- node[above] {\scriptsize $1-p$} (d1);
   \draw[->,line width = 1pt] (s0) -- node[above] {\scriptsize $p$} (d1);
   \draw[->,line width = 1pt] (s1) -- node[above] {\scriptsize $p$} (d0);

 \node at (4,-0.5) (source) {$X_n$};
  \node[sum, right of = source, node distance = 1.5 cm] (plus) {$+$};
  \node[right of = plus, node distance = 1.5cm] (dest) {$Y_n$};
  \node[above of = plus, node distance = 1cm] (noise) {$Z_n$ $\sim \mathrm{Bern}(p)$ };
  \draw[->,line width = 1pt] (source) --  (plus);
  \draw[->,line width = 1pt] (plus) -- (dest);
  \draw[->,line width = 1pt] (noise) --  (plus);
 \end{tikzpicture}
\caption{ Equivalent descriptions of the channel $P_{Y|X}$. }
\label{fig:BSC}
%\vspace{-0.5 cm}
\end{figure}
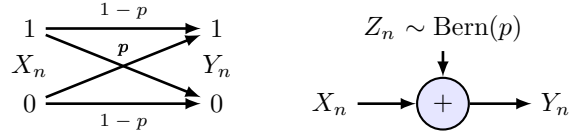

\subsection{Indirect DRF of a binary i.i.d process \label{subsec:problem_statement} }

We now specialize our study of the iRDF to the case where $X^n$ is an i.i.d binary process, $Y^n$ is obtained by passing $X^n$ through a memoryless Binary Symmetric Channel (BSC)
 and for Hamming distortion measure.
%
%Although simple, this model provides many relevant intuitions and has many practical applications.

More specifically, we focus on the case where 
,$X_i \perp X_j, \ i \neq j$ and
\[
Y^n = X^n \oplus Z^n,
\]
where $X^n$ and $Z^n$ are two Bernoulli i.i.d process, independent of each other, with $\mathbb P (X_i = 1) = \al$  and $\mathbb P (Z_i = 1) = p, \ \forall \ i \in \{0\ldots n\}$ respectively.
Accordingly, $\Xcal=\Ycal=\{0,1\}$ and  $Y_i$ is a binary i.i.d process with
\[
\be \triangleq \mathbb P(Y_i = 1) = p \star \al,  \quad \forall \ i\in\{1 \ldots n \}.
\]
For the fidelity criterion at the receiver we consider the case $\widehat{\Xcal}=\{0,1\}$ and %the  criterion $d(x_i,\xh_i)$ be defined as
\ea{
d(x_i,\xh_i)=x_i \oplus \xh_i,
}
which corresponds to the usual Hamming distance between $x^n$ and $\xh^n$.

%Note that $Y^n$ corresponds to the output of the BSC in Figure~\ref{fig:BSC} when the input is $X^n$ and thus the factorization of  \eqref{eq:memoryless} holds
%and the iRDF can be evaluated using \eqref{eq:memoryless RDF}.%
%%
%This provides a great simplification in the closed form calculation of the iRDF using the result of Th. \ref{thm:gallager}.
%In the following we shall investigate the iRDF

%Note that we can always assume that $p<1/2$, since otherwise the encoder can inverse the process $Y$ and make the channel from $X$ to $Y$ into a $\mathrm{BSC}(1-p)$. From the same reason we can also assume that $\al \leq 1/2$. Note that both of these assumptions are possible due to the prior knowledge of $p$ and $\al$ at the encoder and decoder. \\
\begin{rem}
\label{rem: al less 1/2}
Given the symmetry in the source $X_i$ and the noisy observations $Y_i$, we can consider $\al,p\leq 1/2$: the remaining cases can be obtained by complementing the observations $Y^n$ and/or the reconstructions $\Xh^n$.
\end{rem}
In view of Remark~\ref{rem: al less 1/2} we will assume $\alpha,p \leq 1/2$ in the remainder of the paper. 

%\begin{IEEEproof}
%The proof is provided in App. \ref{app:less 1/2}
%\end{IEEEproof}
%For the reason above we can define $\al_m \triangleq  \min\{\al,1-\al\}$, $p_m\triangleq \min\{ p,1-p \}$ and $\be_m \triangleq  (1-\al_m)p_m+\al_m(1-p_m)$.
%
%Given the observation in Rem. \ref{rem: al less 1/2}  and for the sake of brevity we will assume $\alpha,p \leq 1/2$ in the remainder of the paper.

%
\section{Results}
\label{sec:Results}
\subsection{Preliminaries}
From the definition of the iRDF  we can infer some properties of $R_{X|Y}(D)$ for the model in Fig. \ref{fig:BSC}:
%its definition, we can conclude the following facts about $R_{X|Y}(D)$:
\begin{prop}  \label{prop:elementary}
The function $R_{X|Y}(D)$ must satisfy the following properties:
\begin{enumerate}
\item[(i)] $R_{X|Y}(D) = 0$ for any $D\geq \al$.
\item[(ii)] $R_{X|Y}(D)$ is only defined in the interval $D\geq \min\{p,\al\}$.
\item[(iii)] $R_{X|Y}(D)$ is non-decreasing in $p$.
\item[(iv)] $R_{X|Y}(D) \geq R_X(D)$ for any $D$, where
\begin{equation} \label{eq:drf_iid}
R_X(D) = \begin{cases} h(\al)-h(D),& 0\leq D \leq \al,\\ 0, & D > \al.
\end{cases}
\end{equation}
is the RDF of $X$ under the Hamming distortion (see e.g. \cite{ThomasCover}) and corresponds to the case $Y^n=X^n$.
\end{enumerate}
\end{prop}
%
%\begin{IEEEproof}
%(i) follows since $\al$ is the average distortion achieved by the trivial reconstruction $\xh_n= 0$, $n=1,2,\ldots$. (ii) if $p< \al$, then even if the process $Y$ is perfectly available at the decoder, i.e. $R=1$, it is impossible to achieve average distortion less than the crossover probability $p$. (iii) as $p$ gets closer to $1/2$, less information is available on $X$ at the decoder. More precisely, under the assumptions $\al,p \leq 1/2$ we have that $\be \leq 1/2$ and is increasing in $p$. That increases the entropy of the process $Y$ and increases $R_Y(D)=R_{X|Y}(D)$. (iv) $R_X(D)$ corresponds to the case $p=0$, and therefore follows from (iii).
%\end{IEEEproof}

Using the results in Section~\ref{sec:pre}, we can equate the indirect RDF $R_{X|Y}(D)$ to the (direct) RDF $R_{Y}(D)$ by defining the amended  distortion measure $\dh(\cdot,\cdot)$
in \eqref{eq:d_bar_def} obtained as
%
%As explained in , $R_{X|Y}(D) = R_Y(D)$, where $R_Y(D)$ is the direct RDF of $Y$ under the distortion measure $\dh$. We now turn to compute the RDF of $Y$.
%
%For $n=1,2,\ldots$ we have
\begin{align}
\dh_n(y^n,\xh^n)
%& =  \sum_{x^n} d(x^n,\xh^n) \mathbb P\left(X^n=x^n|Y^n=y^n \right) \nonumber \\
& =  \sum_{x^n}  \sum_{i=1}^n \left( x_i \oplus \xh_i \right) \mathbb P\left(X^n=x^n|Y^n=y^n \right),\nonumber  \\
 &  = \sum_{i=1}^n \sum_{x_i \in \{0,1\}} \left( x_i \oplus \xh_i \right) \mathbb P\left(X_i=x_i|Y_i=y_i \right) \nonumber \\
& = \sum_{i=1}^n \mathbb P\left(X_i \neq \hat{x_i} |Y_i=y_i \right)
= \sum_{i=1}^n \dh(y_i,\xh_i).
\label{eq:d_bar_binary}
\end{align}
It follows from \eqref{eq:d_bar_binary} that the new distortion measure $\dh(\cdot,\cdot)$ has an intuitive interpretation:
if $\xh_i \in \{0,1\}$ is the estimate of $X_i$ given the symbol $y_i \in \{0,1\} $, then $\dh(y_i,\xh_i)$ is the probability of making an error in this estimation.
Table \ref{tab:distortion} lists all the possible values of $\dh(y_i,\xh_i)$.

\begin{table}
\begin{center}
\begin{tabular}{c|c|c}
 & $y_i=0$ &  $y_i=1$ \\
\hline
$\xh_i=0$ & \begin{tikz} \node{$ \frac{\al p}{\bar{\be}}$}; \end{tikz}  & \begin{tikz} \node{$\frac{\al \bar{p}}{\be}$}; \end{tikz} \\
\hline
$\xh_i=1$ &  \begin{tikz} \node {$\frac{\bar{\al} \bar{p}}{\bar{\be}}$}; \end{tikz} & \begin{tikz} \node{$\frac{\bar{\al} p}{{\be}}$ }; \end{tikz} \\
\end{tabular}
\end{center}
\caption{Possible values of $\dh(y_i,\xh_i)$ in \eqref{eq:d_bar_binary}}
\label{tab:distortion}
\end{table}
%where in the table above we used the notation $\bar{q}=1-q$ for a number $q\in [0,1]$. \\

\subsection{Main Result}
The next step is to use Theorem~\ref{thm:gallager} to derive $R_{X|Y}(D)$.
%{\color{red} is there a reason why you mixed $\rho$ and $r$ ? (I removed the $\rho$)}
\begin{thm} \label{thm:main_result}
Let
\begin{align}
g( r ) & \triangleq  r\left(D-p\right)+ \log \left( 1-e^{- r (u+v) } \right) \label{eq:g_r} \\
& \quad -\bar{\be} \log\left( 1-e^{-r u} \right)- \be \log \left(1-e^{-r v} \right). \nonumber
\end{align}
The iRDF $R_{X|Y}(D)$ is given by
\[
R_{X|Y}(D) = \begin{cases}
h(\be)-g( r ^\star) & p \leq D \leq \al, \\
0 & D> \min\{\al,p\},
\end{cases}
\]
where $r^\star$ is the unique solution to
\begin{align} \label{eq:r_star_def}
\frac{ \bar{\be} u }{e^{r^\star u}-1}+\frac{ \be v }{e^{r^\star v}-1}  - \frac{u+v }{e^{r^\star (u+v)  }-1} = D-p,
\end{align}
with $u\triangleq (\al-p)/\be$ and $v\triangleq (\bar{\al} - p)\bar{\be}$.
\end{thm}

\begin{IEEEproof}
Only an outline of the proof is provided here: the full proof is provided in App. \ref{app:main_result}.
In view of Proposition~\ref{prop:elementary} it is enough to consider the case $p < D < \al \leq 1/2$.
Assume that equality holds in \eqref{eq:gallager_fk}, then
\[
\begin{pmatrix}
e^{-r\frac{\al p}{\bar{\be}}} & e^{-r \frac{\al \bar{p}}{\be}} \\
e^{-r \frac{\bar{\al}\bar{p}}{\bar{\be}}} & e^{-r \frac{\bar{\al} p}{\be}}
\end{pmatrix}
 \begin{pmatrix} f_0 \\ f_1  \end{pmatrix} = \begin{pmatrix} 1 \\ 1   \end{pmatrix} ,
\]
which implies
\begin{align*}
f_0 & = \frac{1-e^{-r u} }{e^{-r\frac{p \al}{\bar{\be}}}
\left( 1-e^{-r (u + v)  } \right) }, \\
f_1 & = \frac{1-e^{-r v} } {e^{-r \frac{ p \bar{\al}}{\be} } \left( 1-e^{-r (u+v) } \right) },
\end{align*}
where $u \triangleq (\al-p)/\be$ and $v \triangleq (\bar{\al}-p)/\bar{\be}$. Note that both $u$ and $v$ are positive in the domain of interest.
%\begin{align*}
%f_0 & = \frac{e^{-r \frac{ \bar{\al} p}{\be} } - e^{-r \frac{ \al \bar{p}}{\be} }  } {e^{-rp \left( \frac{\al} {\bar{\be}} + \frac{\bar{\al}} {\be}  \right) } -e^{-r\bar{p} \left( \frac{\bar{\al}} {\bar{\be}} + \frac{\al} {\be} \right) } }  \\
%& = \frac{1-e^{-r \frac{\al-p}{\be}} }{e^{-r\frac{p \al}{\bar{\be}}}
%\left( 1-e^{-r \left(\frac{\al-p}{\be} -  \frac{\al-\bar{p}}{\bar{\be}} \right) } \right) }. \\
%f_1 & = \frac{e^{-r \frac{ {\al} p}{ \bar{\be}} } - e^{-r \frac{ \bar{\al} \bar{p}} {\bar{\be} } } } {e^{-rp \left( \frac{\al} {\bar{\be}} + \frac{\bar{\al}} {\be}  \right) } -e^{-r\bar{p} \left( \frac{\bar{\al}} {\bar{\be}} + \frac{\al} {\be} \right) } } \\
%& = \frac{1-e^{-r \frac{\bar{\al}-p}{\bar{\be}}} } {e^{-r \frac{ p \bar{\al}}{\be} } \left( 1-e^{-r \left(\frac{\al-p}{\be} -  \frac{\al-\bar{p}}{\bar{\be}} \right) } \right) }.
%\end{align*}
We next write
\begin{align}
R_{X|Y} (D) & \geq h(\be) + \bar{\be} \log\left( 1-e^{-r u} \right) + \be \log \left(1-e^{-r v} \right) \nonumber \\
&  \quad  - \log \left( 1-e^{- r (u+v) } \right) - r\left(D-p\right) \label{eq:iid_proof_to_maximize}  \\
& = h(\be) - g( r ). \nonumber
\end{align}
In order to maximize the RHS of \eqref{eq:iid_proof_to_maximize}, we take the derivative of $g( r )$ which gives
\begin{align} \label{eq:iid_proof_to_simplify}
g'( r ) = -(D-p) + \frac{u+v }{e^{r^\star (u+v)  }-1} -\frac{ \bar{\be} u }{e^{r^\star u}-1}-\frac{ \be v }{e^{r^\star v}-1}.
\end{align}
It can be shown that $\lim_{r\rightarrow \infty} g'( r ) = p-D < 0$, $\lim_{r\rightarrow 0^+} g'( r ) = 1/2-D > 0$
 and that $g'( r )$ is non-decreasing for $ r >0$.
 All this implies that the maximum of $g( r )$ is obtained at a single point $r^\star$ in the domain $ r >0$ which corresponds
 to $g'( r ^\star)=0$. We conclude that this $r^\star$ maximizes the RHS of \eqref{eq:iid_proof_to_maximize}. \\
It is shown in Appendix~\ref{app:main_result} that for $p < \al \leq 1/2$ and $r= r^\star$, there exist positive
$w_0$ and $w_1$ that satisfy \eqref{eq:gallager_w}.
This implies that substituting $r^\star$ in \eqref{eq:iid_proof_to_maximize} leads to equality, i.e., the iRDF is given by the RHS of \eqref{eq:iid_proof_to_maximize}.
%
%It is shown in Appendix~\ref{app:ws} that for $p < \al \leq 1/2$ and $r= r^\star$, there exists positive $w_0$ and $w_1$ that satisfy \eqref{eq:gallager_w}. This implies that substituting $r^\star$ in \eqref{eq:iid_proof_to_maximize} leads to equality, i.e., the iRDF is given by the RHS of \eqref{eq:iid_proof_to_maximize}.
\end{IEEEproof}
In the special case where $\al = 1/2$ and $p<\al$, we have that $\be = 1/2$ and \eqref{eq:r_star_def} reduces to
\begin{align} \label{eq:iid_proof_half}
\frac{ (\bar{p}-p) }{e^{ r (\bar{p}-p)}-1}
 - \frac{2(\bar{p}-p) }{e^{2 r (\bar{p}-p)   }-1} = D-p,
\end{align}
which leads to
\[
r^\star = \frac{\log\left(\frac{\bar{p}-D}{D-p} \right)}{\bar{p}-p}.
\]
Substituting $r^\star$ in \eqref{eq:g_r} results in $g( r ^\star) = h(\Delta)$,  where
\[
\Delta \triangleq \Delta(D,p) \triangleq  \frac{D-p}{\bar{p}-p}.
\]
It follows from Theorem~\ref{thm:main_result} that
\begin{equation} \label{eq:alpha_half}
R_{X|Y}(D) = \begin{cases}
\log(2)-h\left ( \Delta \right),& p<D<1/2,\\
0, & D\geq 1/2,
\end{cases}
\end{equation}
which is equivalent to \cite[Exc. 3.8]{berger1971rate}. Equation \eqref{eq:alpha_half} has a similar form as the direct RDF \eqref{eq:drf_iid} of a binary i.i.d symmetric process. It is interesting to compare \eqref{eq:alpha_half} to \eqref{eq:drf_iid} and to observe how the properties of $R_{X|Y}$ anticipated in Proposition~\ref{prop:elementary} are expressed in the special case of \eqref{eq:alpha_half}.
\begin{enumerate}
\item[(i)] $D=1/2$ corresponds to $h(\Delta)=h(1/2)=\log(2)$.
\item[(ii)] The domain of $R_{X|Y}(D)$ is $0\leq \Delta$ or $p\leq D$.
\item[(iii)] $\Delta$ is decreasing in $p$ and therefore $R_{X|Y}(D)$ is increasing in $p$.
\item[(iv)] \eqref{eq:alpha_half}  reduces to \eqref{eq:drf_iid} for $p=0$.
\end{enumerate}

\begin{figure}
\begin{center}
\begin{tikzpicture}
\node at (0,0) {\includegraphics[trim=0cm 0cm 0cm 0cm,  ,clip=true,scale=0.45]{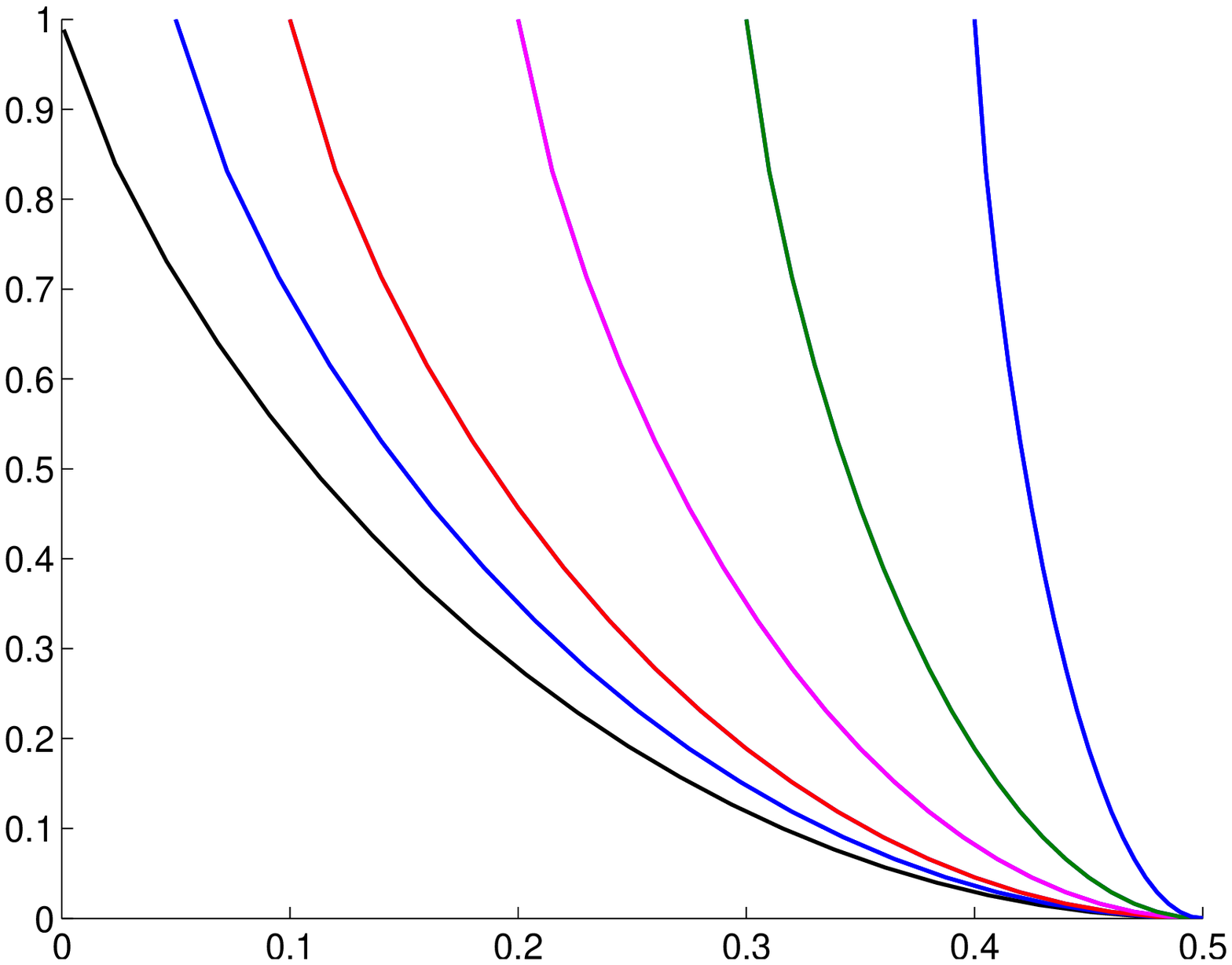}};
\node[rotate=90] at (-4.2,0.3) {\small {$R~[bits]$ }} ;
\node at (0,-3.2) {\small $D$};

\draw[dashed] (-2.75,-2.8) node[below] {\tiny $0.05$} -- (-2.75,2.7);
\draw[dashed] (-2.05,-2.8) -- (-2.05,2.7);
\draw[dashed] (-0.6,-2.8) -- (-0.6,2.7);
\draw[dashed] (0.8,-2.8) -- (0.8,2.7);
\draw[dashed] (2.2,-2.8) -- (2.2,2.7);
%\node at (-3.5,2) {\scriptsize $p=0.01$};

\end{tikzpicture}
\caption{\label{fig:various_p} $R_{X|Y}(D)$ for $\al = 1/2$ and various values of $0\leq p<1/2$ that correspond to the vertical dashed lines.}
\end{center}
\vspace{-.5 cm}
\end{figure}

The slope of  $R_{X|Y}(D)$ is an important parameter since it determines the maximal  return in code-rate reduction
for each additional distortion unit the system can tolerate. In the range $p \leq D \leq 1/2$, this slope is given by
\begin{equation} \label{eq:slope}
 \frac{1}{\po-p} \log \left(\frac{\po-D}{D-p}\right).
\end{equation}
Note that this slope is more steep than the slope of $R_X(D)$, and goes to infinity as $p$ approaches $1/2$ (see Fig.~\ref{fig:various_p}).
This fact confirms the intuition that an increment in the bit-rate when describing noisy measurements is less effective in reducing
 distortion as the intensity of the noise increases. \par
Another interesting factor is the rate at which $R_{X|Y}(D)$ changes with $p$ for a fixed $p \leq D \leq \al \leq 1/2$.
This rate represents the amount of excess coding needed as a result of increasing uncertainty on the source in order to keep a fixed distortion. \\

\begin{figure}
\begin{center}
\begin{tikzpicture}
\node at (0,0) {\includegraphics[trim=0cm 0cm 0cm 0cm,  ,clip=true,scale=0.45]{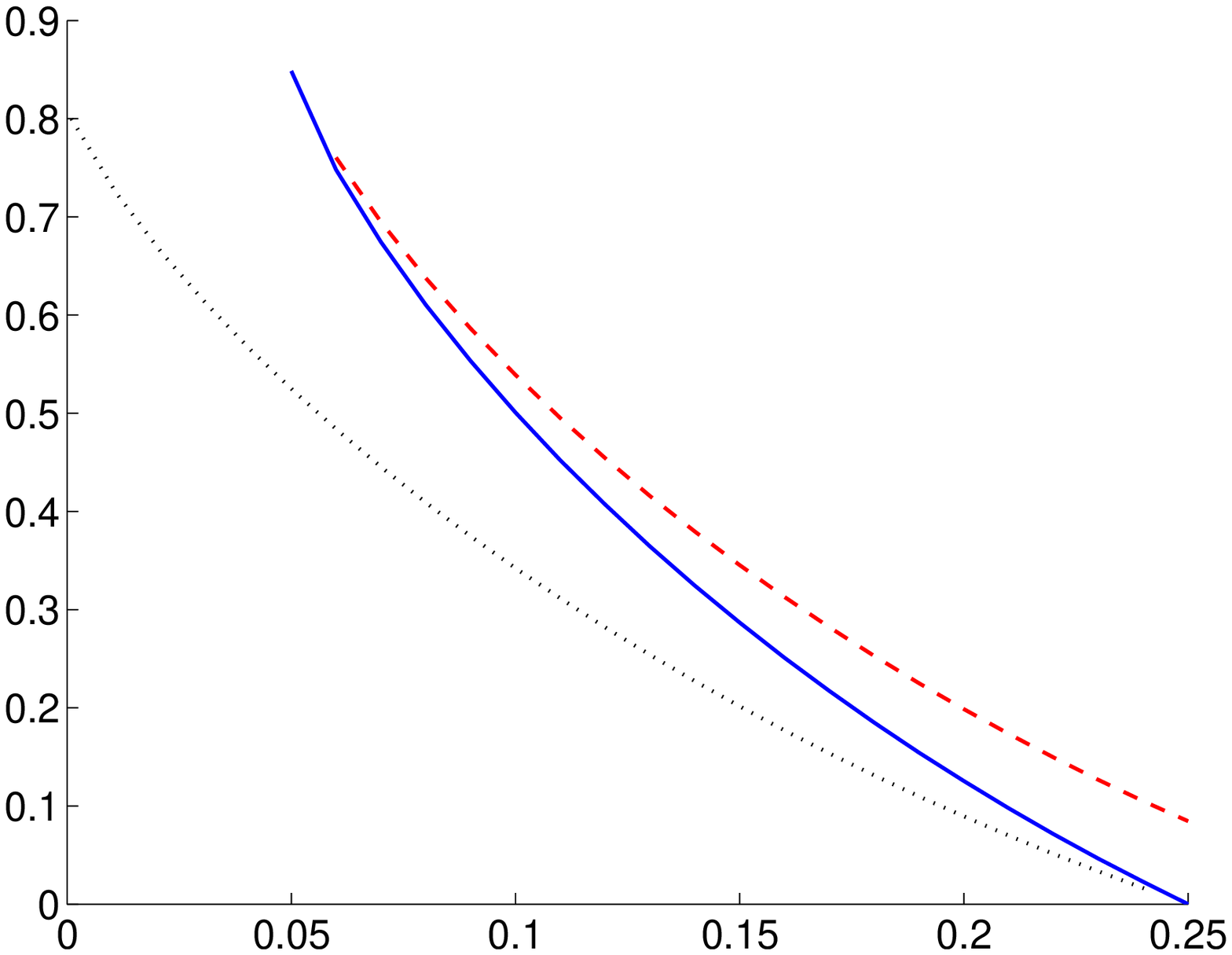}};
\node[rotate=90] at (-4.2,0.3) {\small {$R~[bits]$ }} ;
\node at (0,-3.2) {\small $D$};
\draw[dashed] (-2.05,-2.8) -- (-2.05,2.4);
\node[rotate = -48] at (-0.7,0) {\color{blue} \small $R_{X|Y}(D)$};
\node[rotate = -39] at (0.5,0) {\color{red} \small $h(\be)-h(\Delta)$};
\node[rotate = -37] at (-1.2,-0.5) {\small $h(\al)-h(D)$};
\draw[dashed] (3.5,-2.7) --(0.6,-0.48);
%\node at (-3.5,2) {\scriptsize $p=0.01$};

\end{tikzpicture}
\caption{\label{fig:iRDF_iid_bound} $R_{X|Y}(D)$, $R_X(D)$ and the upper bound \eqref{eq:bound} for $\al = 1/4$ and $p=0.05$.  }
\end{center}
\vspace{-.5 cm}
\end{figure}

Due to the similarity between \eqref{eq:alpha_half} and \eqref{eq:drf_iid}, it may be tempting to guess that $R_{X|Y}(D)$ is given in a similar form
to \eqref{eq:alpha_half} even in the case where $\al < 1/2$.
While an exact solution of \eqref{eq:r_star_def} is hard to obtain in general,
it is possible to obtain the following bound.
\begin{thm} \label{thm:bound}
For any $p,\al \in [0,1]$ and $D\geq p$,
\begin{equation}  \label{eq:bound}
R_{X|Y}(D) \leq  h(\be)-h\left ( \Delta \right),
\end{equation}
where $\Delta = (D-p)/(1-2p)$.
\end{thm}
\begin{IEEEproof}
The proof is provided in App. \ref{app:bound}.
%It is enough to assume that $p \leq \al \leq 1/2$. For $\al=1/2$ we have
%\begin{equation} \label{eq:proof_bound_half}
%g( r ) =  \log \left(1-e^{- r (\po-p)}\right) - \log \left(1-e^{-2 r (\po-p)}\right).
%\end{equation}
%We will show that for all $ r >0$, the difference between $g( r )$ that corresponds to any $p\leq D \leq 1/2$ and the one that corresponds to $\al=1/2$ is always positive.
%This difference can be written as
%\begin{align} \label{eq:proof_bound_diff}
%&\delta( r ) \triangleq \beb \log \left( \frac{1-e^{- r (u+v)} } {\left(1-e^{-ru} \right) \left(1+e^{- r (\po-p) } \right) } \right) \nonumber \\
%&  \quad \quad \quad \quad + \be \log \left( \frac{1-e^{- r (u+v)} } {\left(1-e^{-rv} \right) \left(1+e^{- r (\po-p)} \right) }  \right).
%\end{align}
%The result follows by noting that $\lim_{r\rightarrow \infty}\delta( r )=0$ and the derivative of $\delta( r )$ is strictly positive for any $ r >0$.
\end{IEEEproof}

The bound in Theorem~\ref{thm:bound} is illustrated in Figure~\ref{fig:iRDF_iid_bound}. The fact that \eqref{eq:bound} is not tight in general can be easily seen since $\Delta > \be$ at $D=\al$ for $\al \neq 1/2$. In fact, due to the convexity of $R_{X|Y}(D)$, a better bound can be obtained by adding the point $R_{X|Y}(\al)=0$ to the bounding curve and taking the convex closure, as illustrated by the dashed line in Figure~\ref{fig:iRDF_iid_bound}. \par
In view of Theorems \ref{thm:gallager}, \ref{thm:main_result} and \ref{thm:bound}, the results in this paper can be summarized by the following statement. For $p < D < \al$ and any $ r >0$ we have
\begin{equation} \label{eq:summary}
h(\be)- g( r ) \leq R_{X|Y}(D) \leq h(\be) - h(\Delta),
\end{equation}
where the LHS holds with equality if and only if $r$ satisfies \eqref{eq:r_star_def}, and the RHS holds with equality if and only if  $\al = 1/2$.

%%%%%%%%%%%%%%%%%%%%%%%%%%%%%%%%%%%%%%%%%%%%%%%%%%%%%%%%%%%%%%%%%%%%%%%%%%%%%%%%

\section{Conclusions}
\label{sec:Conclusions}
This paper  studies the indirect rate-distortion problem for a binary i.i.d.  source under the Hamming distortion given its noisy observation through a binary symmetric channel.
The indirect rate distortion problem is an extension of the rate distortion problem in which the encoder is provided with a noisy observation of the source sequence.
%
%The indirect rate distortion problem can actually be formulated as a classic rate distortion problem by choosing a distortion measure which accounts for the uncertainty on the source realization introduced by the noisy observations.
%
We investigate the rate-distortion tradeoff for the simple scenario of a binary source, bit flipping noise and Hamming distortion.
Although conceptually simple, this model provides a number of key intuitions on more general models and illustrates important tradeoffs for practical systems. % such as centralized sensor networks.
For instance, by deriving the relationship between rate and distortion at each noise level, we make it possible to determine how the sampling error and the communication error probabilities can be balanced in a remote sensor to obtain a desired target end-to-end quality of measurement.
%it is more effective to reduce the noise in the measurements than
%scenario of a binary source, noisy observations obtained
%
%
%\appendices
%

\appendices

%\subsection{Proof of Rem. \ref{rem: al less 1/2}}\label{app:less 1/2}
%\textcolor[rgb]{1.00,0.00,0.00}{complete}

\section{\label{app:main_result}}

In this Appendix we complete the proof of Theorem~\ref{thm:main_result} by showing the existence of positive $w_0$ and $w_1$ that satisfy \eqref{eq:gallager_w}. \\

From the expression to $f_0$ and $f_1$ we obtain:
\begin{align} \label{eq:iid_proof_w0}
w_0 &= \bar{\beta}  \frac{ e^{-r \frac{\bar{\alpha}p}{\beta}}}{e^{-r \frac{\bar{\alpha} p}{\beta}}-e^{-r \frac{{\alpha} \bar{p}}{\beta}} } - \beta  \frac{ e^{-r \frac{{\alpha}\bar{p}}{\beta}}}{e^{-r \frac{{\alpha} p}{\bar{\beta}}}-e^{-r \frac{\bar{\alpha} \bar{p}}{\bar{\beta}}} }, \\
w_1 &=\beta  \frac{ e^{-r \frac{{\alpha}{p}}{\bar{\beta}}}}{e^{-r \frac{{\alpha} p}{\bar{\beta}}}-e^{-r \frac{\bar{\alpha} \bar{p}}{\bar{\beta}}} } -  \bar{\beta}  \frac{ e^{-r \frac{\bar{\alpha}\bar{p}}{\bar{\beta}}}}{e^{-r \frac{\bar{\alpha} p}{\beta}}-e^{-r \frac{{\alpha} \bar{p}}{\beta}} }, \label{eq:iid_proof_w1}
\end{align}
We need to show that \eqref{eq:iid_proof_w0} and \eqref{eq:iid_proof_w1} are positive for any $p < \alpha < 1/2$ and $r= r^\star$. The case where $\alpha = 1/2$ were treated above and leads to $w_0 = w_1 = 1/2$. If $p=\alpha$, then it follows from Proposition~\ref{prop:elementary} that $R_{X|Y}(D)$ is defined only for $D \geq \alpha$ and equals zero. We will therefore assume $p \leq D \leq  \alpha < 1/2$. Another way to write \eqref{eq:iid_proof_w0} and \eqref{eq:iid_proof_w1} is
\begin{align*}
w_0 (r) &= \frac{\beb} {1-e^{-ru }}- \frac{\be e^{-r \frac{\alb \al }{\be \beb} (\po-p) }} { 1-e^{-r v}  }. \\
w_1(r) & =\frac{ \be}{1-e^{-r  v}} -\frac{\beb e^{-r \frac{\alb \al }{\be \beb} (\po-p) } }{1-e^{-r u}}.
\end{align*}
Since $u>0$, $v>0$ and $\po-p>0$ in the domain of interest, it can be shown that $\lim_{r\rightarrow \infty} w_0 (r)= \bar{\beta}$ and that the derivative of $w_0 (r)$ is negative for any $r>0$. This implies that $w_0(r)>0$ for all values of $r$ in the domain of interest and in particular at $r=r^\star$.\\

For $w_1$ we can show that $\lim_{r\rightarrow 0^+}w_1(r) = -\infty$, $\lim_{r\rightarrow \infty} w_1 (r) = \beta$ and it is monotonically increasing for $r>0$. By continuity of $w_1(r)$, it follows that there exists $r_0>0$ with $w_1(r_0)=0$ such that $w_1(r)<0$ whenever $r<r_0$ and $w_1(r)>0$ whenever $r>r_0$.  Since we have seen in the proof of Theorem~\ref{thm:main_result} that $g'(r)$ has similar behavior with a unique root $r^\star$, we conclude that if $g'(r_0)<0$, then $r^\star  > r_0$ and then $w_1(r^\star)>0$. It is therefore enough to show that $g'(r_0)<0$. Indeed, at $r=r_0$ we have
\[
\frac{ \be}{1-e^{-r  v}} = \frac{\beb e^{-r \frac{\alb \al }{\be \beb} (\po-p) } }{1-e^{-r u}}.
\]
Substituting that in the expression for $g'(r)$ we obtain
\begin{align*}
b(r) \triangleq g'(r=r_0) = -D+p+\frac{u+v}{e^{r(u+v)}-1} - \frac{\bar{\beta}u}{e^{ru}-1} -  \frac{\beb v e^{-r \left(\frac{\alb \al }{\be \beb} (\po-p) -u\right) } }{e^{r u}-1}.
\end{align*}
Define
\[
a(r) = -D+p+\frac{u+v}{e^{r(u+v)}-1} - \frac{\bar{\beta}u}{e^{ru}-1} -  \frac{\beb v } {e^{r u}-1}.
\]
Since
\[
\frac{\al \alb} {\be \beb}(\po-p)-u > 0,
\]
we have that $a(r) > g'(r=r_0)$ for all $r>0$. In addition, $\lim_{r\rightarrow \infty} a(r) = -D+p <0$ and
\[
a'(r) = (u+v) \left( -\frac{u+v}{\left(e^{r(u+v)}-1 \right)^2} + \frac{u \beb} {\left(e^{ru}-1 \right)^2} \right),
\]
which is positive for all $r>0$. We conclude that $b(r)<a(r)<0$ for all $r>0$. This proves the claim.

%We also know that  $g'(r^\star)=0$, $\lim_{r\rightarrow} g'(r) = p-D<0$, $\lim_{r\rightarrow 0^+} g'(r) = 1/2-D >0$ and that $g'(r)$ is monotonically increasing for all $r>0$.

\section{}
\label{app:bound}

\subsection*{Proof of Th. \ref{thm:bound}}
It is enough to assume that $p \leq \al \leq 1/2$. For $\al=1/2$ we have
\begin{equation} \label{eq:proof_bound_half}
g( r ) =  \log \left(1-e^{- r (\po-p)}\right) - \log \left(1-e^{-2 r (\po-p)}\right).
\end{equation}
We will show that for all $ r >0$, the difference between $g( r )$ that corresponds to any $p\leq D \leq 1/2$ and the one that corresponds to $\al=1/2$ is always positive.
This difference can be written as
\begin{align} \label{eq:proof_bound_diff}
&\delta( r ) \triangleq \beb \log \left( \frac{1-e^{- r (u+v)} } {\left(1-e^{-ru} \right) \left(1+e^{- r (\po-p) } \right) } \right) \nonumber \\
&  \quad \quad \quad \quad + \be \log \left( \frac{1-e^{- r (u+v)} } {\left(1-e^{-rv} \right) \left(1+e^{- r (\po-p)} \right) }  \right).
\end{align}
The result follows by noting that $\lim_{r\rightarrow \infty}\delta( r )=0$ and the derivative of $\delta( r )$ is strictly positive for any $ r >0$.

\bibliographystyle{IEEEtran}
\bibliography{IEEEabrv,markov}

\end{document}